\documentclass[prb,twocolumn,showpacs,superscriptaddress,floatfix]{revtex4-1}
\usepackage{graphicx}
\usepackage{amssymb,amsmath}
\usepackage{hyperref}

\usepackage{color}

\begin{document}

\title{Spinon confinement: dynamics of weakly coupled Hubbard chains}

\author{Marcin Raczkowski}
\affiliation{Department of Physics and Arnold Sommerfeld Center for Theoretical Physics, 
             Ludwig-Maximilians-Universit\"at M\"unchen, D-80333 M\"unchen, Germany}
\affiliation{Institut f\"ur Theoretische Physik und Astrophysik,
             Universit\"at W\"urzburg, Am Hubland, D-97074 W\"urzburg, Germany}
\author{Fakher F. Assaad}
\affiliation{Institut f\"ur Theoretische Physik und Astrophysik,
             Universit\"at W\"urzburg, Am Hubland, D-97074 W\"urzburg, Germany}

\date{\today}

\begin{abstract}
Using large-scale determinant quantum Monte Carlo simulations
in combination with the stochastic analytical continuation, 
we study two-particle dynamical correlation functions in the anisotropic square lattice of 
weakly coupled one-dimensional (1D) Hubbard chains at half-filling and in the presence of weak frustration. 
The evolution of the static spin structure factor upon increasing the interchain coupling
is suggestive of the transition from the power-law decay of  spin-spin correlations in the 1D limit
to long-range  antiferromagnetic order  in the quasi-1D regime and at $T=0$. 
In the numerically accessible regime of interchain couplings, the charge sector remains  
gapped.  The low-energy momentum dependence of the spin excitations   
is well described by the linear spin-wave theory with the largest intensity located around 
the antiferromagnetic wave vector.  This magnon mode corresponds to a bound state of two spinons.  
At higher  energies the spinons deconfine  and we  observe  signatures of the two-spinon continuum  which 
progressively fade away as a function  of interchain hopping.

\end{abstract}

\maketitle

\section{Introduction}
In the realm of the solid state, controlling dimensionality implies that the thermal energy (or frequency) 
is larger that the effective coupling which triggers a dimensional crossover.~\cite{Kivelson00}  
Under those circumstances,  there is no coherence 
between  the lower-dimensional units such that they effectively decouple. 
The dimensional crossover is  particularly interesting  when  elementary excitations fractionalize in 
the  lower-dimensional limit.~\cite{Giamarchi_book} 
For example, neutron scattering experiments on weakly coupled spin ladders of CaCu$_2$O$_3$
show the two-spinon continuum at frequencies  
larger than the interchain exchange and their \emph{confinement} in the higher-dimensional  ladder 
system emerging  at lower energies.~\cite{Lake10} 
Another experimental realization is provided by BaCu$_2$Si$_2$O$_7$ and KCuF$_3$  which  
consist of weakly coupled spin-1/2 chains.~\cite{Zhe00,Zhe01,Lake05}
At high frequencies one observes  the two-spinon continuum,  signalizing free spinons. 
In the low-frequency limit, pairs of spinons bind to form the Goldstone mode (spin-waves) 
of the broken-symmetry phase. 

In addition to gapless \emph{transverse} spin-wave excitations, nearly disordered by quantum fluctuations 
quasi-1D antiferromagnets are expected to exhibit anomalous (i.e., beyond the standard spin-wave theory) 
\emph{longitudinal} mode at finite energy corresponding to fluctuations in the staggered magnetic moment.
The presence of longitudinal spin excitations near the disordered transition has been predicted
within the quantum Sine-Gordon field theory for weakly coupled to form an anisotropic \emph{cubic} 
lattice spin-1/2 chains treating the interchain couplings as perturbation.~\cite{Schultz96,Essler97}
The longitudinal mode has been resolved in KCuF$_3$,~\cite{Lake00,Lake05prb} 
while so far only a broad continuum feature has been found in BaCu$_2$Si$_2$O$_7$.~\cite{Zhe02,Zhe03} 

The dimensional crossover is not limited to spin systems but also plays the essential role 
in our understanding of Bechgaard  salts. 
In these organic compounds a dimensional-crossover-driven insulator to metal Mott transition 
has been observed.~\cite{Vescoli98} A complete theoretical description  of this phenomenon is still 
missing and continues to capture interest.~\cite{Bier01,Ess02,Tsu07, Rib11,Mou11,Raczkowski12} 

The aim of this paper is to study numerically the change in the nature of low-energy excitations on 
coupling one-dimensional (1D) half-filled  Hubbard chains.  
In contrast to previous work,~\cite{Raczkowski12} our aim is to study two-particle quantities: 
spin and charge  dynamical structure factors. 
Since the 1D regime is dominated by strong momentum dependence 
of the self-energy, an accurate evaluation of the two-particle spectra requires the necessity to 
include vertex corrections.  Hence, one faces a challenging task going beyond the dynamical cluster 
approximation schemes.~\cite{Maier05,Kot06}
    
To handle the full complexity of the problem, we have adopted the finite-temperature auxiliary-field 
quantum Monte Carlo (QMC) algorithm.~\cite{BSS} It allows us to carry out simulations on lattice sizes 
ranging up to $20\times 20$ in the presence of {\it weak}  frustration and  {\it close}  to the 1D limit. 
Here, the limiting factor is the onset of the negative sign problem which ultimately leads to an 
exponential scaling  of the computational time as a function of system size and inverse  
temperature $\beta=1/T$. We restrict our studies to the spin rotationally symmetric case and address 
the emergence of \emph{transverse} spin-wave mode in a dimensional crossover from 1D to 
quasi-two-dimensional (2D) systems.

Our main results  can be summarized as follows.  1D half-filled Hubbard chains are insulating due to 
the relevance of umklapp scattering.
Below the charge gap, spin dynamics is characterized by the two-spinon continuum  
and absence of long-range magnetic order.  Within our resolution, we shall see that  this state 
is unstable towards interchain  hopping, since down to our smallest considered values, the ground state 
develops long-range  antiferromagnetic (AF) order.  
This result is consistent with the one put forward in Ref.~\onlinecite{Affleck94} and confirmed 
numerically in Refs.~\onlinecite{Sandvik99,Kim00} albeit in the absence of frustration and in the 
realm of quantum spin systems.   
This implies that for any value of  interchain hopping, there is an energy scale~\cite{Sandvik11}  
below which one will observe a crossover  in the dynamical spin structure factor from a \emph{two-spinon} 
continuum to a \emph{spin-wave}  mode.  Our real-frequency QMC data, extracted by carrying 
out the stochastic analytic continuation~\cite{Beach04a} of the imaginary time-dependent 
spin-spin correlation function, provide ample support for this interpretation. In addition, 
our results indicate strong thermal effects in quantum magnets with weakly confined 
spinons.~\cite{Vojta09} 

The paper is organized as follows. Section~\ref{Model} defines the model Hamiltonians. 
Section~\ref{simple} briefly describes approximate dynamical response of the spin system 
in two limiting cases: 
(i)  the  two-spinon continuum  typical of the 1D spin chain can be accounted for  
within a SU(2) spin symmetric mean-field approximation, and  
(ii) spin dynamics of the 2D magnetically ordered phase at $T=0$ 
is conveniently described within the spin-wave theory.~\cite{Anderson52,Kubo52}   
Section~\ref{Results} presents our main results, i.e, the dimensional-crossover-driven 
evolution of static as well as dynamical properties of the ground state. 
The conclusions are summarized in Sec.~\ref{Conclusions}. 
 
\section{Models} 
\label{Model}

We study the Hubbard model on a strongly anisotropic square lattice at half-filling,
\begin{equation}
H=-\sum_{\pmb{ij},\sigma}t^{}_{\pmb{ij}}
   c^{\dag}_{{\pmb i}\sigma}c^{}_{{\pmb j}\sigma} +
   U\sum_{\pmb i}n^{}_{{\pmb i}\uparrow}n^{}_{{\pmb i}\downarrow} 
   -\mu\sum_{\pmb i,\sigma}n_{{\pmb i}\sigma},
\label{Hubb}
\end{equation}
where the electron hopping $t_{\pmb{ij}}$ is $t$ ($t_{\perp}$) on the intrachain (interchain)
bonds, $\mu$ is the chemical potential and we set $U/t=3$. Due to  perfect nesting 
of the Fermi surface, the Hubbard model~(\ref{Hubb}) is expected to develop N\'eel order in the 
presence of any infinitesimally small interchain coupling in the $T=0$ 
limit.~\cite{Affleck94,Sandvik99,Kim00}  
Here, we allow for a finite \emph{diagonal} next-nearest neighbor hopping
$t'=-t_{\perp}/4$.  It reduces  nesting  and makes the scenario of immediate magnetic 
ordering less obvious.~\cite{Affleck96} 

Additionally, we consider the spin $S=1/2$ Heisenberg model with nearest neighbor interactions 
$J$ ($J_{\perp}$) along the intrachain (interchain) bonds, respectively, extended by next-nearest 
neighbor interaction $J'$:
\begin{equation}
H_{J}=
J\sum_{\langle\pmb{ij}\rangle_{\parallel}} {\pmb S}_{\pmb i}\cdot{\pmb S}_{\pmb j}
  +J_{\perp}\sum_{\langle\pmb{ij}\rangle_{\perp}} {\pmb S}_{\pmb i}\cdot{\pmb S}_{\pmb j}
  +J'\sum_{\langle\langle\pmb{ij}\rangle\rangle } {\pmb S}_{\pmb i}\cdot{\pmb S}_{\pmb j}.
\label{HJ}
\end{equation}
The interaction $J'$ competes with the tendency toward long-range AF order and gives rise 
to geometric frustration. We shall discuss in Sec.~\ref{Dynamic} to what extent spin-wave 
dispersion in the Heisenberg model describes the low-energy QMC spin excitations of the 
Hubbard model.

\section{Approximate spin dynamics}
\label{simple}

In this Section, we discuss two limiting cases of the spin dynamics: 
(i) two-spinon continuum typical in one-dimension, and 
(ii) spin-wave modes characteristic of a higher-dimensional magnetically ordered phase. 
The overall spectral features of the two limits can be reproduced using simple but different
 approximation schemes.

\subsection{Two-spinon continuum in one-dimension}
\label{MF}

In the presence of any finite Hubbard interaction, the relevance of umklapp scattering in the 
1D regime and  at  half-filling, opens a gap for charge excitations while leaving the spin 
sector gapless.  
Below the charge gap,  one can model the relevant physics by a spin-only $S=1/2$ Heisenberg chain. 
In  the 1D limit, the  magnon excitation which  one produces by flipping a
spin  decomposes into  two spinons corresponding to  domains walls  in the AF background.  
Hence, the  spin dynamics is characterized by  the two-spinon  continuum.

\begin{figure}[t!]
\begin{center}
\includegraphics[width=0.43\textwidth]{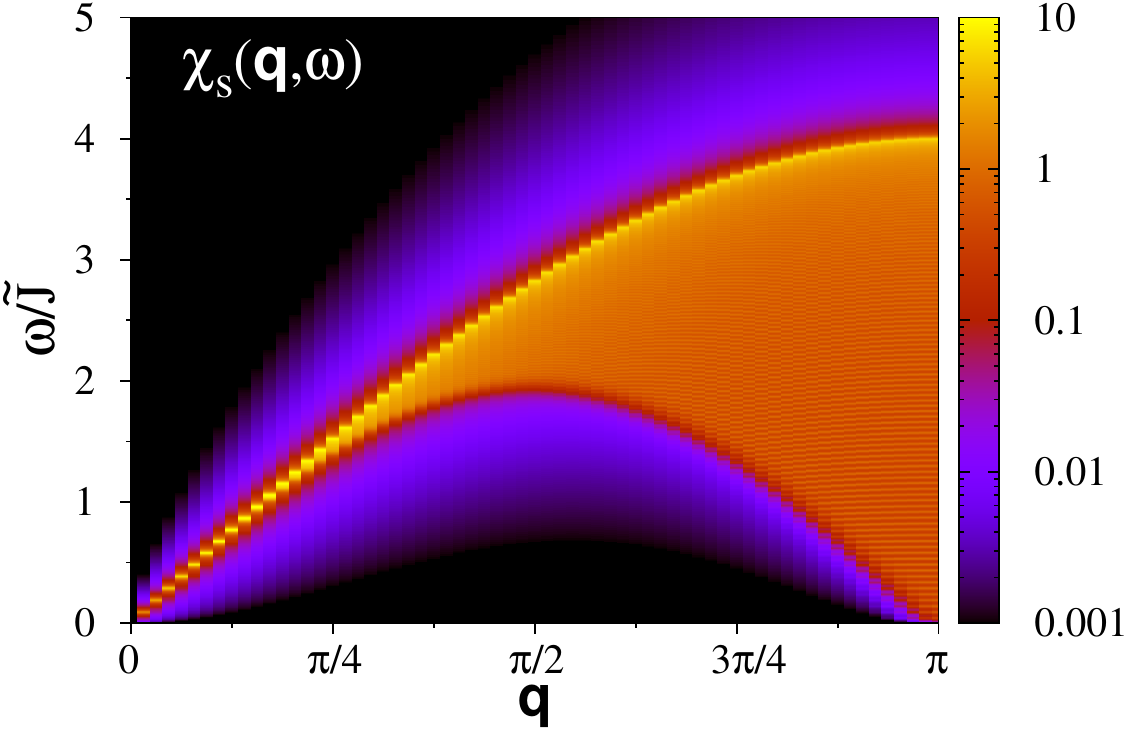}
\end{center}
\caption
{(Color online)
Two-spinon continuum  of the spin susceptibility $\chi_s(\pmb{q}, \omega)$ Eq.~(\ref{chi})
as obtained from  the SU(2) spin symmetric MF approximation of the 1D Heisenberg model.
}
\label{spinon}
\end{figure}

A simple understanding of this phenomenon is provided by a mean-field (MF) decoupling 
which conserves the SU(2) spin symmetry.~\cite{Affleck88} 
Starting from the 1D Heisenberg model, one can adopt a fermion representation of the spin-operator 
$  {\pmb S}_{\pmb i} =  \frac{1}{2} \sum_{\sigma,\sigma'} f^{\dagger}_{{\pmb i},\sigma} 
\vec{ \sigma}_{\sigma,\sigma'} f^{}_{{\pmb i},\sigma'} $  subject to the constraint,
\begin{equation} 
 \sum_{\sigma} f^{\dagger}_{{\pmb i}\sigma} f^{}_{{\pmb i}\sigma} = 1.  
\label{slave}
\end{equation}
Next, using the relation:  
\begin{equation}
{\pmb S}_{\pmb i}  \cdot {\pmb S}_{\pmb j} - \frac{1}{4} = 
- \frac{1}{4}\bigl(  D^{\dagger}_{\pmb i \pmb j} D^{}_{\pmb i \pmb j}  
           + D^{}_{\pmb i \pmb j} D^{\dagger}_{\pmb i \pmb j}  \bigr), 
\end{equation}
with 
$  D_{\pmb i \pmb j} = \sum_{\sigma}   f^{\dagger}_{{\pmb i},\sigma} f^{}_{{\pmb j},\sigma} $ 
and introducing the bond-order parameter $ \chi = \langle   D_{\pmb i \pmb j} \rangle $ 
yields  the following MF Hamiltonian,
\begin{equation}
   H^{\rm{MF}}_{\widetilde{J}}   =  -\widetilde{J}    
\sum_{\langle {\pmb i} {\pmb j} \rangle_{\parallel},\sigma }  
\bigl( f^{\dagger}_{{\pmb i}, \sigma}  f^{}_{{\pmb j}, \sigma}  
+   f^{\dagger}_{{\pmb j}, \sigma}  f^{}_{{\pmb i}, \sigma}   \bigr),
\label{H_MF}  
\end{equation}
with a renormalized coupling $\widetilde{J}\equiv J \chi/2$.
At the  MF level, the constraint Eq.~(\ref{slave}) is satisfied only on average, i.e.,  
$ \sum_{\sigma}  \langle f^{\dagger}_{{\pmb i},\sigma} f^{}_{{\pmb i},\sigma}\rangle  = 1$,  
such that the spinon Fermi wave vector reads $k_{\rm F}= \pi/2$. Finally, 
the corresponding  spin susceptibility in the $z$-direction is given by,
\begin{equation}
\chi_s(\pmb{q}, \omega)    =  \frac{1}{4N} \sum_{{\pmb k}}  
\frac{f_{\rm F}(\pmb{k}+\pmb{q} ) -f_{\rm F}(\pmb{k}) } 
{\varepsilon({\pmb{k})} - \varepsilon({\pmb{k} + \pmb{q}) - \omega - i \delta} },
\label{chi}
\end{equation}
where $f_{\rm F}(\pmb{k})$ is the Fermi function and $\varepsilon({\pmb{k})}$ is the 1D tight-binding 
dispersion of the MF Hamiltonian Eq.~(\ref{H_MF}).  
The intensity plot of $\chi_s(\pmb{q}, \omega)$  is shown in Fig.~\ref{spinon}. It depicts the 
two-spinon continuum with  two gapless excitations  at ${\pmb q}=0$ and $ {\pmb q}=2k_{\rm F} = \pi$.

\subsection{Linear spin-wave theory} 
\label{LSWT}

In spatial dimensions greater than one, the spinons  bind  and form magnons, 
which are nothing but the Goldstone modes of the continuous SU(2) spin broken symmetry.    
The magnons are well accounted within the linear spin-wave theory (LSWT) in the leading 
$1/S$ order.~\cite{Anderson52,Kubo52}
The spin-wave expansion is valid about the ground state with a finite classical order parameter.
In contrast, there is no long-range order in one-dimension and the magnons decay into pairs of spinons.
The signature of spin fractalization is then seen as continuum of excited states. 
Clearly, the dimensional crossover must affect the nature of elementary spin excitations. 
In particular, when spinons bind, the continuum of excitations in the dynamical spin structure factor 
is expected to give way to well defined  sharp magnon peaks described by the LSWT.

Below we discuss the LSWT magnon dispersion as well as quantum corrections to the magnetic order 
parameter $\langle S^z\rangle$. We focus on the AF phase within the anisotropic Heisenberg 
model defined in Eq.~(\ref{HJ}). 
The starting point is the Holstein-Primakoff transformation expressing the spin operators 
in terms of bosonic operators $\{a_i^{\dagger},a_i^{}\}$,
\begin{equation}
S_i^z=S-a_i^{\dagger}a_i^{}, \hskip .5cm
S_i^+\simeq \sqrt{2S}a_i^{}, \hskip .5cm
S_i^-\simeq \sqrt{2S}a_i^{\dagger}. 
\label{hp}
\end{equation}
The effective bosonic problem is then solved by employing
subsequent Fourier and Bogoliubov transformations. The latter 
diagonalizes a $2\times 2$ dynamical matrix at each momentum {\pmb q}.  
The corresponding magnon dispersion reads,
\begin{equation}
\omega_{\pmb q}=2S\sqrt{ \xi^2_{\pmb q} -\gamma^2_{\pmb q} },
\label{magnon}
\end{equation}
with 2D structure factors,
\begin{align}
\xi_{\pmb q} &= J+J_{\perp}-2J'(1-\cos q_{x}\cos q_{y}),\\
\gamma_{\pmb q} &= J\cos q_{x}+J_{\perp}\cos q_{y}.
\end{align}

\begin{figure}[t!]
\begin{center}
\includegraphics[width=0.43\textwidth]{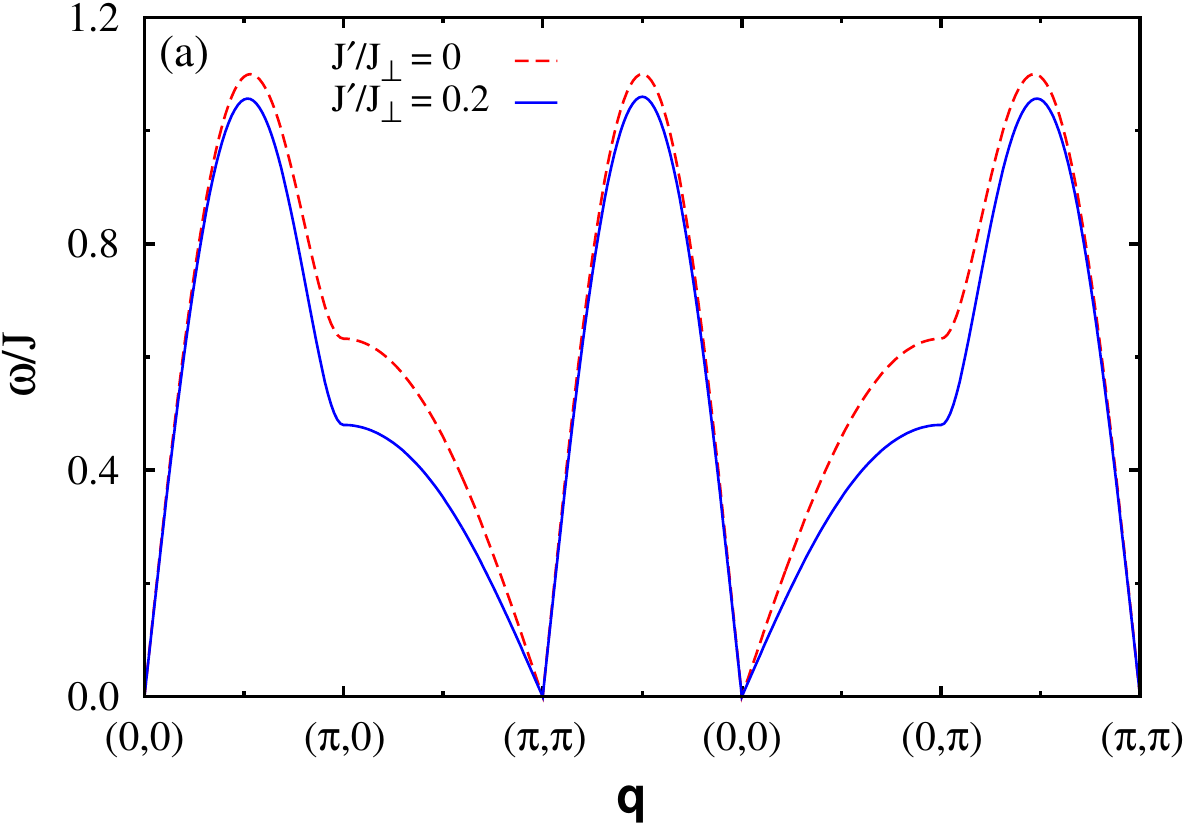}\\
\includegraphics[width=0.43\textwidth]{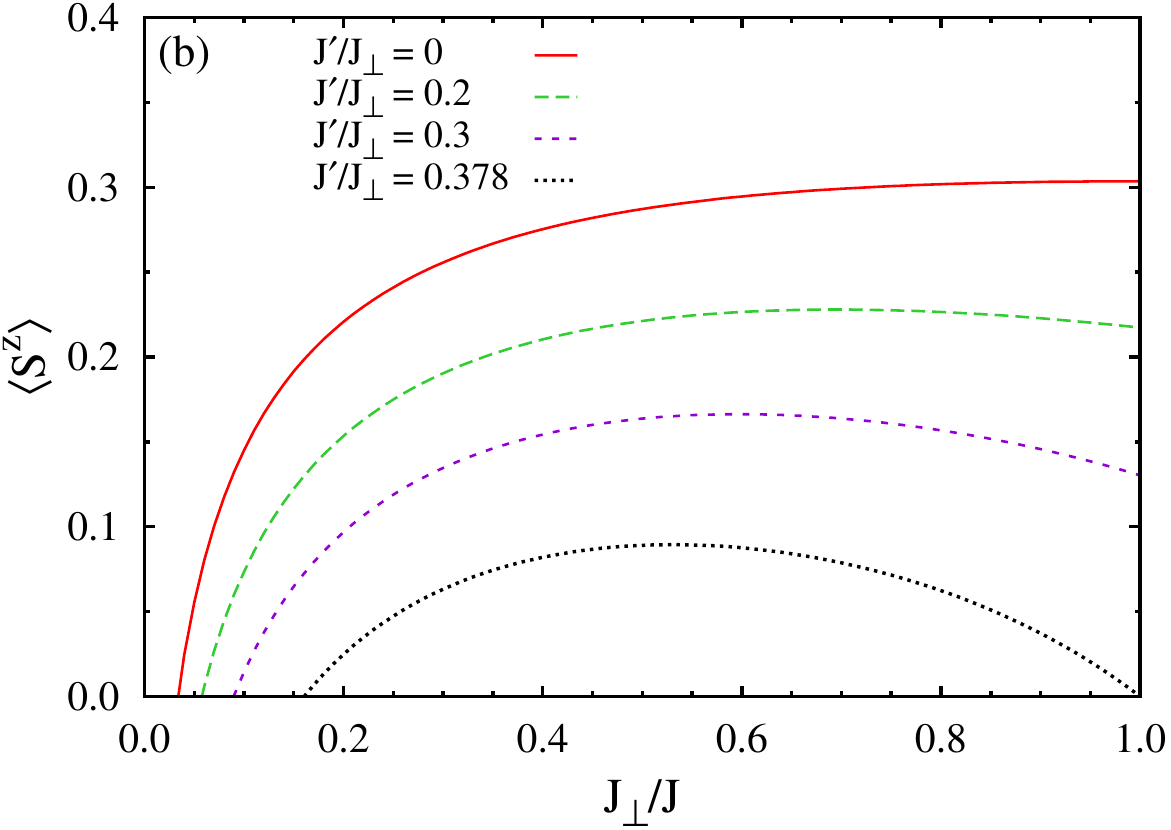}
\end{center}
\caption
{(Color online)
(a) Spin-wave dispersion in the AF phase within the anisotropic $S=1/2$ Heisenberg model (\ref{HJ})
with $J_{\perp}/J=0.1$;
(b) Effective magnetic order parameter Eq.~(\ref{Sz}) within the LSWT  as a function of increasing 
interchain interaction $J_{\perp}$.
 }
\label{spin_wave}
\end{figure}

Spin-wave dispersion of the Heisenberg model (\ref{HJ})  with $J'=0$ is displayed for a 
representative value of $J_{\perp}/J=0.1$  in Fig.~\ref{spin_wave}(a). 
The low-energy part of the spectrum is given by gapless Goldstone modes with a linear momentum 
dependence,
\begin{equation} 
\omega_{\alpha}({\pmb q})\simeq D_{\alpha} {{\pmb q}},
\label{gold}
\end{equation}
with $D_{\parallel}$ ($D_{\perp}$) being the spin-wave stiffness along (perpendicular to) 
the chains, respectively.  We compare it with the dispersion obtained with a finite frustrated 
interaction $J'$. One observes in Fig.~\ref{spin_wave}(a) that even small 
$J'/J_{\perp}=0.2$ modifies the magnon dispersion. In particular, the frustration manifests 
itself as: 
(i) flattening of the magnon band around momenta ${\pmb q}=(\pi,0)$ and ${\pmb q}=(0,\pi)$, and 
(ii) reduction of the stiffness constant $D_{\perp}$ leaving $D_{\parallel}$ almost intact.

It is instructive to consider the effect of quantum fluctuations on the 
classical order parameter $S$ in the presence of frustrated interaction $J'$. 
Using the standard scheme of the LSWT,~\cite{Raczkowski02} one can show that the size 
of quantum corrections in the $T\to 0$ limit is given by,
\begin{equation}
\delta\langle S^z\rangle = S\int\frac{d^2{\pmb q}}{(2\pi)^2}\frac{\xi_{\pmb q}}{\omega_{\pmb q}}-\frac{1}{2}.
\label{moment}
\end{equation}
The effective magnetic order parameter,
\begin{equation}
\langle S^z\rangle=S-\delta\langle S^z\rangle,
\label{Sz}
\end{equation}
for spin $S=1/2$ is plotted as a function of $J_{\perp}$ in Fig.~\ref{spin_wave}(b).
On the one hand, previous QMC studies of weakly coupled $S=1/2$ Heisenberg 
chains,~\cite{Sandvik99,Kim00} reveal that a finite critical value $J_{\perp}^*/J\simeq 0.03$ 
required for the onset of long-range AF order is an artifact of the LSWT. 
The failure follows from the increased importance of higher-order corrections to 
the local order parameter $\langle S^z\rangle$ in the 1D limit.~\cite{Iga05}
On the other hand, the LSWT captures the frustrating role of next-nearest neighbor 
coupling $J'$ and predicts noticeable enhancement of $J_{\perp}^*$. 
We list critical $J_{\perp}^*$ for a few  representative  values of $J'$ in Table~\ref{J_c}. 
Among them, $J'/J_{\perp}=0.378$ is expected within the LSWT to fully suppress long-range AF 
order in the isotropic 2D limit. 
The same tendency has been found in the QMC simulations of the half-filled 
2D Hubbard model when the next-nearest neighbor hopping is sufficiently large.~\cite{Hirsch87}

\begin{table}[t!]
\caption{\label{J_c}
Critical value $J_{\perp}^*$ required to stabilize long-range AF order in the quasi-1D regime in the
presence of frustrated coupling $J'$.
}
\begin{tabular}{ccccccccc}
\hline \hline
 $J'/J_{\perp}$   &&  0       &&   0.2              &&    0.3         &&    0.378  \\
\hline
 $J_{\perp}^*/J $   &&  0.033   &&   0.058            &&    0.09         &&   0.162   \\
\hline \hline
\end{tabular}
\end{table}

\section{Results}
\label{Results}

We proceed now to present our QMC results for the Hubbard model (\ref{Hubb}).  The results were obtained using  a finite-temperature implementation of the auxiliary-field QMC algorithm  
(see Ref.~\onlinecite{Assaad08_rev} and references therein). 
It involves a separation of the one-body kinetic $H_t$ and two-body Hubbard interaction $H_U$ terms 
with the help of the Trotter decomposition,  
\begin{equation}
 e^{-\Delta \tau \left( H_U + H_t \right) } \simeq 
e^{-\Delta \tau H_U } e^{ -\Delta  \tau H_t }. 
\end{equation}
Typically, we have used a finite imaginary time step $\Delta \tau t = 1/6$. 
This introduces an overall  controlled systematic error of order $(\Delta  \tau)^2$. 
Finally, we have opted for a  discrete, Ising,  Hubbard-Stratonovitch  field coupling to 
the $z$-component of the magnetization. With this setup,  we have carried out simulations 
for lattice sizes ranging from $L=8$ to $L=20$ and in the broad range of temperatures $t/5\le T\le t/30$. 
Severe minus-sign problem arising due to a finite next-nearest neighbor hopping $t'=-t_{\perp}/4$  
disabled QMC simulations beyond $t_{\perp}/t=0.35$.  

\subsection{Equal time spin-spin correlation}
\label{Static}

A salient feature of the 1D interacting system is the power-law decay of all correlation 
functions.~\cite{Giamarchi_book}  
To investigate dimensional-crossover-driven effects  in the nature of spin degrees of freedom, 
we calculate Fourier transform of the equal-time spin-spin correlations,
\begin{equation}
S({\pmb q})=\frac{4}{3}
    \sum_{{\pmb r}}e^{i{\pmb q}\cdot{\pmb r}}\langle {\pmb S}({\pmb r})\cdot{\pmb S}({\pmb 0}) \rangle.
\end{equation}

\begin{figure}[t!]
\begin{center}
\includegraphics[width=0.43\textwidth]{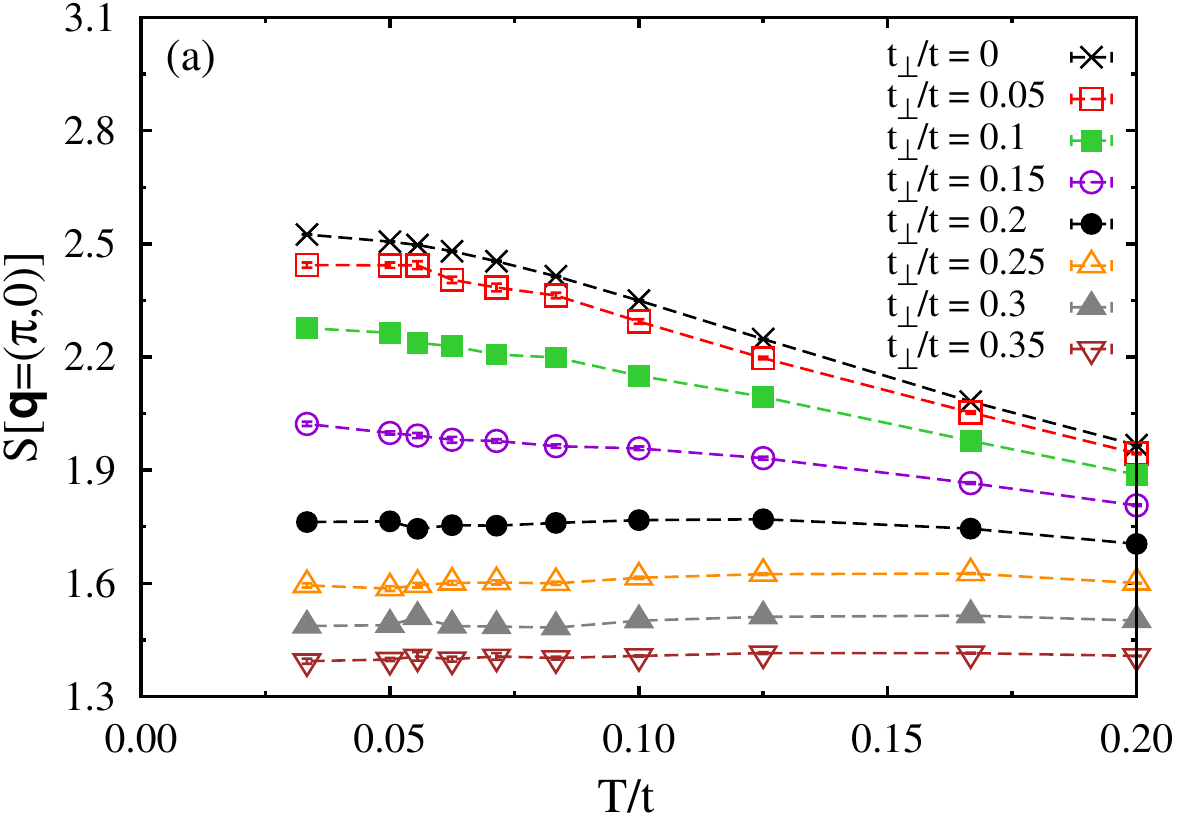}\\
\includegraphics[width=0.43\textwidth]{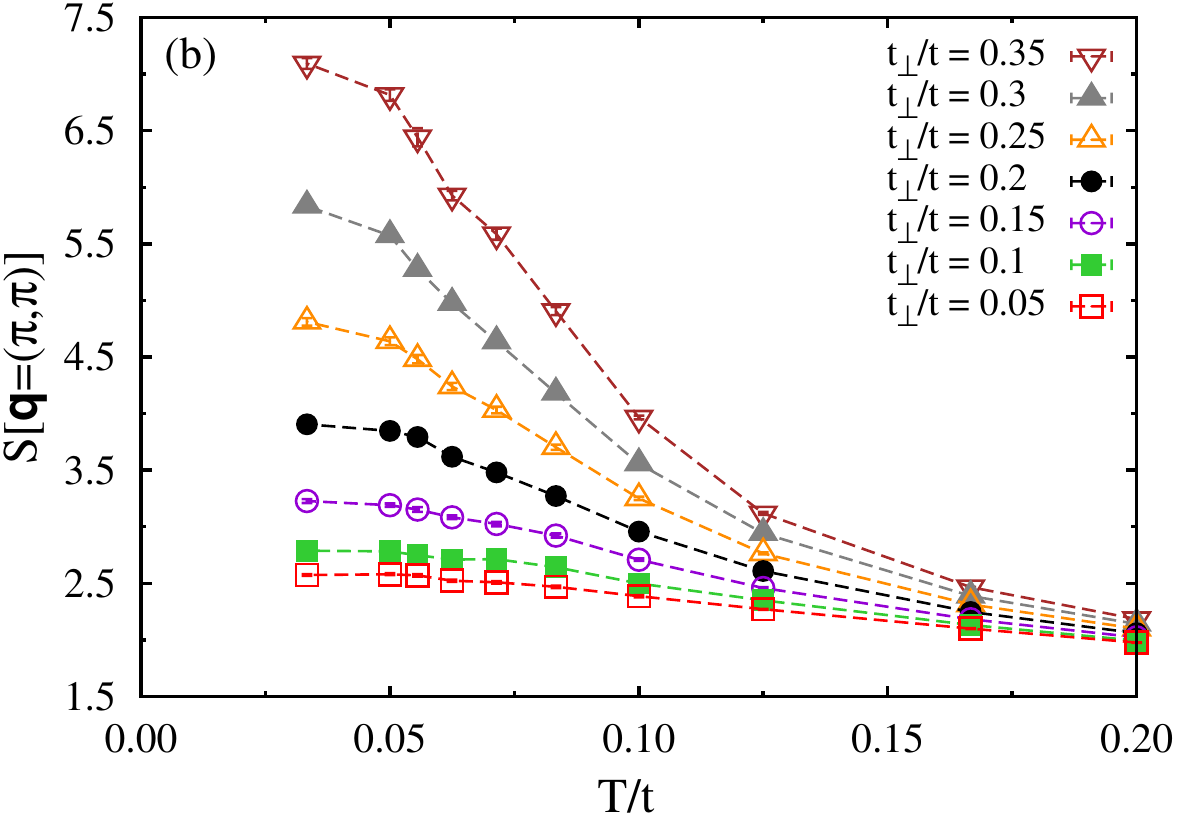}
\end{center}
\caption
{(Color online)
Temperature dependence of the static spin structure factor $S({\pmb q})$
found on the 16$\times 16$ lattice at momentum:
(a) ${\pmb q}=(\pi,0)$ and
(b) ${\pmb q}=(\pi,\pi)$.
}
\label{S_T}
\end{figure}

Figure~\ref{S_T}(a) illustrates temperature dependence of the static spin structure factor $S(\pi,0)$. 
This observable measures spin-spin correlations along the chains. 
One finds that the low-$T$ enhancement of $S(\pi,0)$ in the 1D limit is progressively replaced 
by a flat behavior on increasing $t_{\perp}$. The observed change is consistent with a transition 
from the power-law decay of spin-spin correlations in the isolated chains to an exponential falloff 
in the system of weakly coupled chains.
However, sufficiently high temperature, i.e., larger than warping of the Fermi surface, should 
restore the 1D behavior of the spin-spin correlations. This indeed happens close to $T=t/5$ since 
the value of the spin structure factor $S(\pi,0)$ for $t_{\perp}/t=0.05$ almost matches that of 
the 1D regime, see Fig.~\ref{S_T}(a).

\begin{figure}[t!]
\begin{center}
\includegraphics[width=0.43\textwidth]{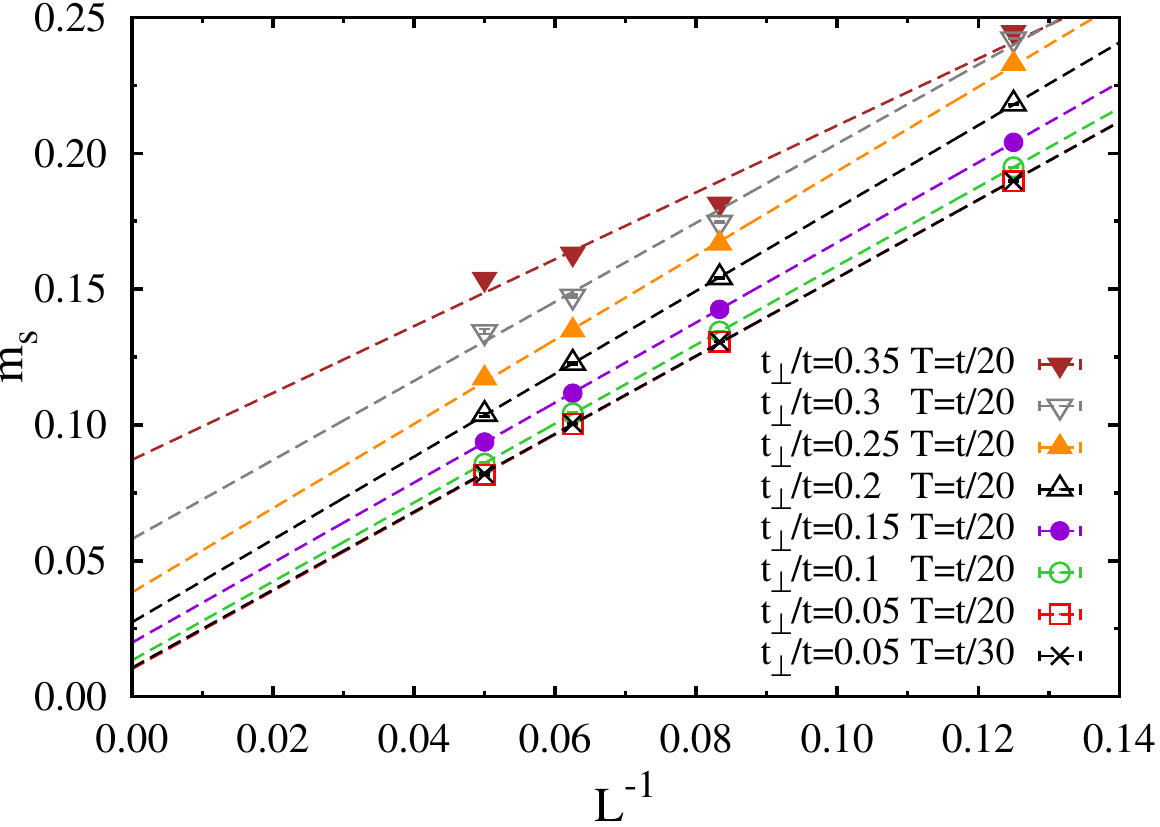}
\end{center}
\caption
{(Color online)
Finite size extrapolation of the staggered magnetic moment $m_s$.
}
\label{S_L}
\end{figure}

The low-$T$ increase of staggered spin structure factor $S(\pi,\pi)$  shown in Fig.~\ref{S_T}(b), 
is suggestive of the transition triggered by $t_{\perp}$ to the AF phase.
To provide further support for the onset of broken-symmetry ground state, we plot in Fig.~\ref{S_L}
finite-size scaling of the staggered magnetic moment,
\begin{equation}
m_s=\lim_{L\to\infty}\sqrt{\frac{S(\pi,\pi)}{L^2}}.
\end{equation}
Down to our smallest considered values $t_{\perp}/t=0.05$ and in the effective zero-temperature limit, 
a finite value of $m_s$ is found in the thermodynamic limit.
Clearly, the weak frustration brought by small $t'=-t_{\perp}/4$ does not prevent the formation
of long-range AF order.
The absence of finite critical interchain coupling is supported by neutron 
scattering data on Sr$_2$CuO$_3$, a quasi-1D $S=1/2$ antiferromagnet with a very small N\'eel temperature
 $T_N/J\simeq 5K$. In this case a \emph{continuous} reduction of the magnetic moment
in the limit of vanishing interchain interactions has been found.~\cite{Shirane97}

\subsection{Spin and charge dynamics}
\label{Dynamic}

Having found the signatures of long-range AF order in the static spin-spin correlation function 
$S(\pi,\pi)$, we proceed to discuss spin $S({\pmb q},\omega)$  and charge  $C({\pmb q},\omega)$  
dynamical structure factors defined as: 
\begin{equation} 
	S({\pmb q},\omega) = \frac{\chi_s''({\pmb q},\omega)}{1-e^{-\beta \omega}} 
\; \; {\rm and } \; \;
       C({\pmb q},\omega) = \frac{\chi_c''({\pmb q},\omega)}{1-e^{-\beta \omega}}. 
\end{equation}
Here,  $\chi_c$ and $\chi_s$ correspond to the generalized  charge and spin susceptibilities. 
The susceptibilities  can be obtained from imaginary-time displaced two-particle correlation functions,
\begin{align}
\langle S^z({\pmb q}, \tau)  S^z(-{\pmb q},0) \rangle  &= 
\frac{1}{\pi} \int {\rm d} w \;  \frac{e^{-\omega\tau}}{1 - e^{-\beta \omega}} \; \chi_s'' ({\pmb q},\omega),  \label{sqw} \\ 
\langle N({\pmb q}, \tau)   N(-{\pmb q},0) \rangle     &= 
\frac{1}{\pi} \int {\rm d} w \; \frac{e^{-\omega\tau}}{ 1 - e^{-\beta \omega} } \;  \chi_c''({\pmb q},\omega) \label{cqw},
\end{align}
where,
\begin{align} 
 S^z({\pmb q})  &= \frac{1}{\sqrt {L} } \sum_{\pmb r} e^{i {\pmb q}\cdot {\pmb r} } 
\left( n_{{\pmb r}\uparrow} - n_{{\pmb r}\downarrow}  \right),  \label{sz}\\ 
  N({\pmb q})  &= \frac{1}{\sqrt {L} } \sum_{\pmb r} e^{i {\pmb q}\cdot {\pmb  r} } 
\left( n_{{\pmb r}\uparrow} + n_{{\pmb r}\downarrow}  - n\right) \label{n}. 
\end{align}
with $n=\sum_{\sigma}\langle n_{{\pmb r},\sigma}\rangle$ being the average filling level. 
We extracted the real-frequency two-particle spectra by analytically continuing the imaginary-time 
QMC data with the use of a stochastic version of the Maximum Entropy method.~\cite{Beach04a} 

The dynamical spin structure factor $S({\pmb q},\omega)$ is related to the static one $S({\pmb q})$ 
through the sum rule:  
\begin{equation}
S({\pmb q}) = \frac{1}{\pi}\int {\rm d} w \; S({\pmb q},\omega).
\label{sum}
\end{equation}
Hence, we shall resolve in $S({\pmb q},\omega)$ redistribution of magnetic spectral weight in 
the thermally-driven  dimensional crossover. The latter was anticipated in Sec.~\ref{Static} from the 
evolution of the static spin-spin correlation function  $S({\pmb q})$, see Fig.~\ref{S_T}.     
However, our primary goal is to elucidate the evolution of the \emph{two-spinon} continuum typical 
of the  $S=1/2$ Heisenberg chain into the \emph{single-magnon} mode known as a low-energy magnetic 
excitation  in a 2D antiferromagnetically ordered phase at 
$T=0$.~\cite{Jarrell92,Sandvik01,Scalettar02}

\subsubsection{1D limit: deconfined spinons}

\begin{figure}[t!]
\begin{center}
\includegraphics*[width=0.43\textwidth]{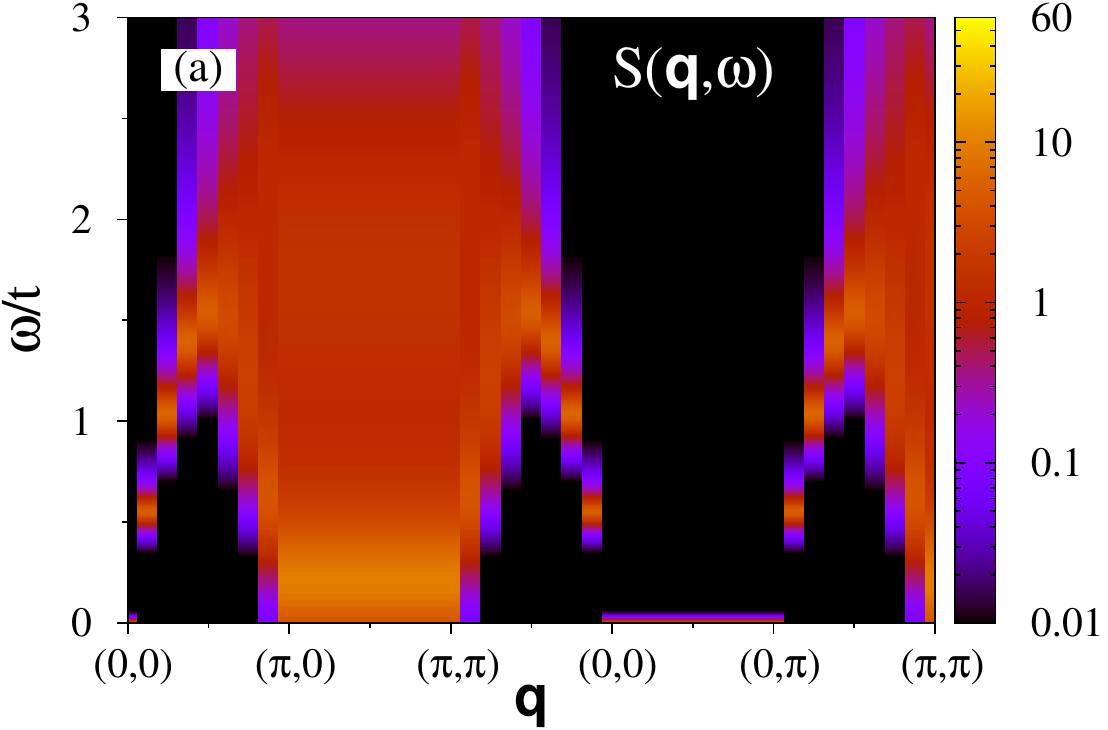}\\
\includegraphics*[width=0.43\textwidth]{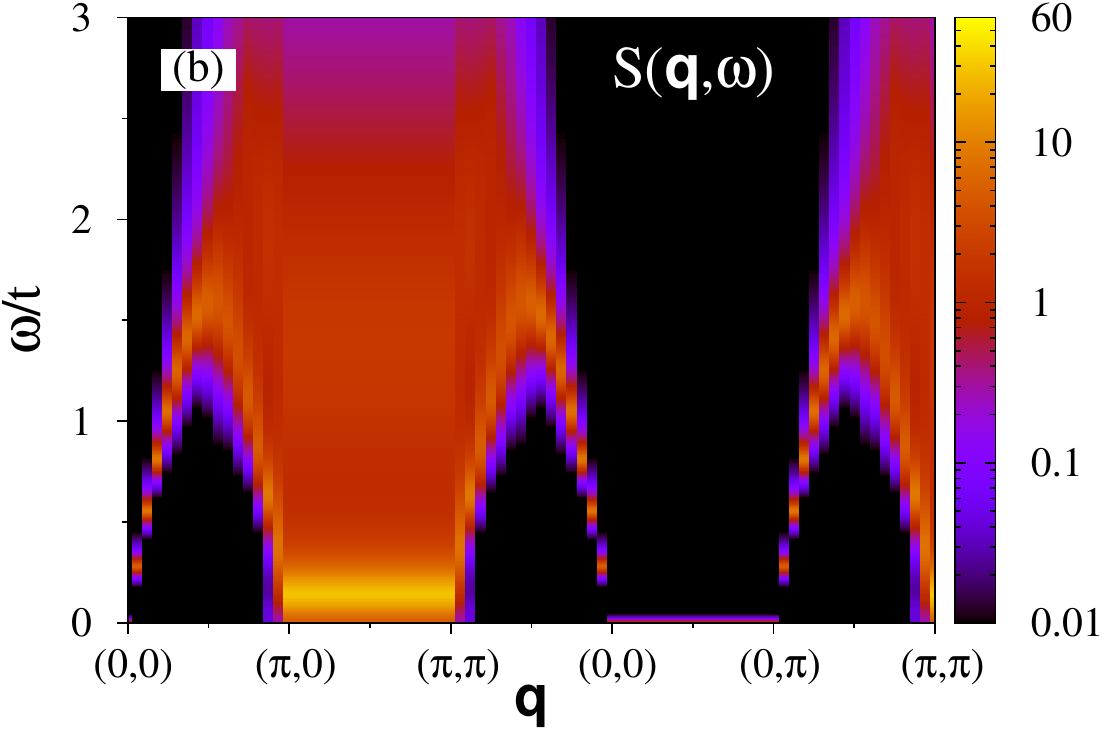}
\end{center}
\caption {(Color online) Dynamical spin structure factor   $S({\pmb q},\omega)$
obtained on: (a) 16-site chain at $T=t/10$ and (b) 32-site chain at $T=t/20$.
}
\label{Sqw0}
\end{figure}

\begin{figure}[t!]
\begin{center}
\includegraphics*[width=0.43\textwidth]{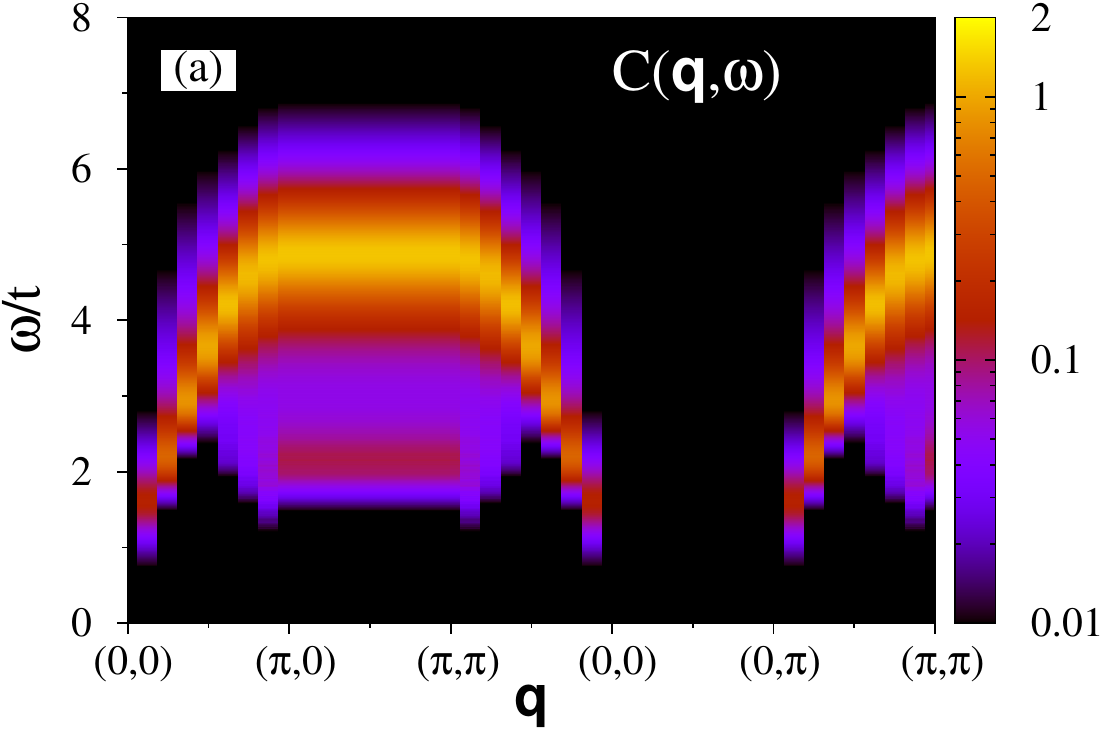}\\
\includegraphics*[width=0.43\textwidth]{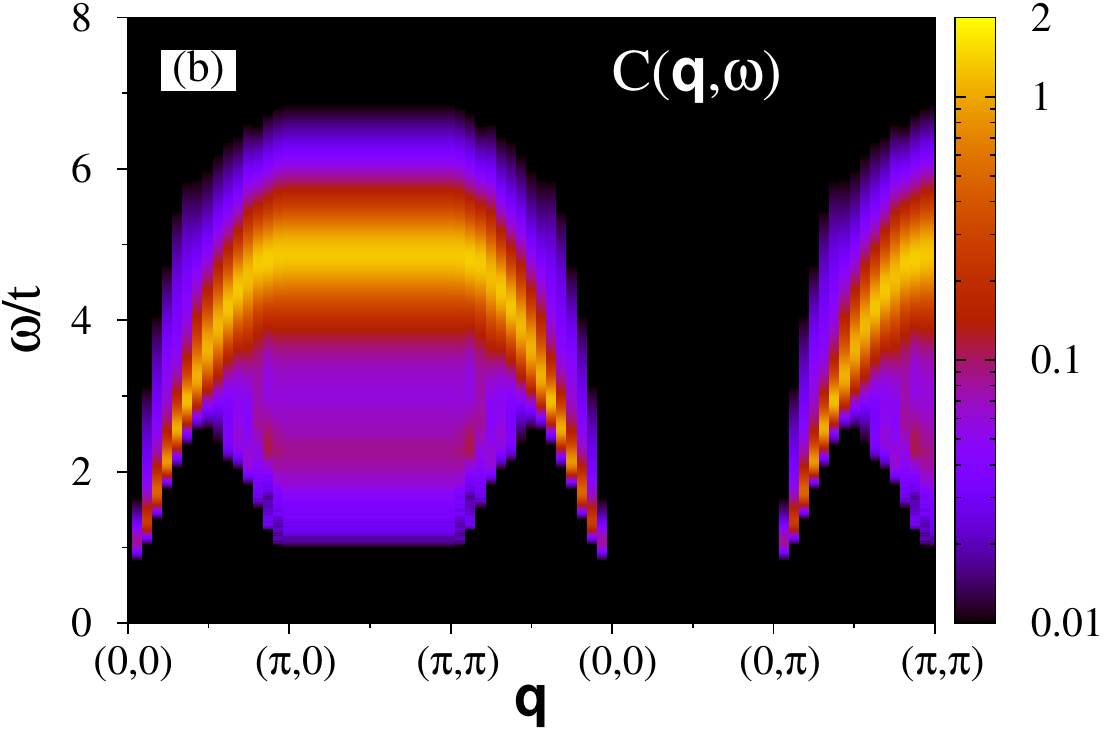}
\end{center}
\caption {(Color online) Dynamical charge structure factor   $C({\pmb q},\omega)$
obtained on: (a) 16-site chain at $T=t/10$ and (b) 32-site chain at $T=t/20$.
}
\label{Cqw0}
\end{figure}

We begin with the system of isolated ($t_{\perp}=0$) half-filled 1D Hubbard chains.
The corresponding intensity plots of the spin and charge excitation spectrum
are shown in Figs.~\ref{Sqw0} and \ref{Cqw0}.  In the realm of bosonization, spin and charge  degrees of
freedom decouple. The energy scale,  as measured from the Fermi energy, up to which this  
picture  survives depends on the strength of the interaction.  For example, in the single-particle 
spectra signatures of spin-charge separation are  detected  up to very high energies for interactions of
the order of the bandwidth.~\cite{Abend06}  Beyond this  energy scale, one would expect
to recover the noninteracting picture  where the spin and charge dynamical structure factors
have \emph{equivalent}  supports.  This regime corresponds to the \emph{particle-hole} continuum.
With this in mind, we can analyze the  data of Figs.~\ref{Sqw0} and \ref{Cqw0}.
At first glance, one will not detect a particle-hole continuum  -- as defined above -- within the plotted 
energy range indicative of dominant role of vertex corrections.  
Hence the data, again in the considered energy range, should be understood in terms of collective spin
and charge excitations.

The dynamic charge structure factor $C({\pmb q},\omega)$  shows features similar to those 
seen in Ref.~\onlinecite{Penc12} using time-dependent density matrix renormalization group.  
The charge sector is gapped with a relatively large  charge gap $\Delta/t>1$ (c.f.~Fig.~\ref{Cqw0}).
Above the charge gap, one can detect aspects of the charge dynamics of Luttinger liquids, 
with low-lying charge excitations located at long wavelengths as well as at 
${\pmb q} = 2 k_{\rm F} = \pi$. Since the charge is fully gapped, the spin dynamics can be 
understood predominantly with a spin-only model which we will approximate by the  $S=1/2$ 
Heisenberg chain.  
Its most prominent feature is the two-spinon continuum of states
bounded from below and above by, ~\cite{Clo62,Yam69}
\begin{equation}
\frac{\pi}{2}J|\sin({\pmb q})|\leq\omega({\pmb q})\leq\pi J|\sin({\pmb q}/2)|. 
\label{CP}
\end{equation}
The upper bound corresponds to two spinons with the same momentum ${\pmb q}/2$.
The lowest-lying excited states come from two-spinons in which one spinon has momentum ${\pmb q}$ and the
other has zero or $\pi$. As shown in Fig.~\ref{Sqw0}(b), these excitations carry the main 
spectral weight at low-$T$ as expected in the strong coupling regime $U/t\gg1$ dominated by virtual 
hopping of electrons.~\cite{Jarrell93,Sandvik97,Essler05,Ben07,Barthel09}
This should be contrasted with the MF spectrum in Fig.~\ref{spinon}. 
Although the latter reproduces the overall form of $S({\pmb q}, \omega)$, it is biased with respect 
to the distribution of weight since the MF approximation does not capture the criticality of the SU(2) 
spin-chain correctly. In this respect, the MF spectrum is similar to $S({\pmb q}, \omega)$ obtained in 
the weak coupling limit $U/t\ll 1$ with strong electron itinerancy effects.~\cite{Essler05}

The presence of zero-frequency magnetic weight at ${\pmb q}=\pi$ and equivalent momenta shall
give rise to immediate binding of spinons on coupling the 1D chains. 
This conjecture, supported by the analysis of the static spin structure factor $S(\pi,\pi)$ in 
Sec.~\ref{Static}, is suggestive of the emergence of low-energy spin-waves characteristic of 
the magnetically ordered phase. 
In contrast, no  spectral weight in the charge and spin sector is found along the $(0,0)\to(0,\pi)$ 
direction, i.e., perpendicular to the chains. Hence, the dynamics of the dimensional crossover should 
be particularly clearly visible in this part of the Brillouin zone.

\subsubsection{Spinon confinement in the quasi-1D limit}

\begin{figure}[t!]
\begin{center}
\includegraphics*[width=0.43\textwidth]{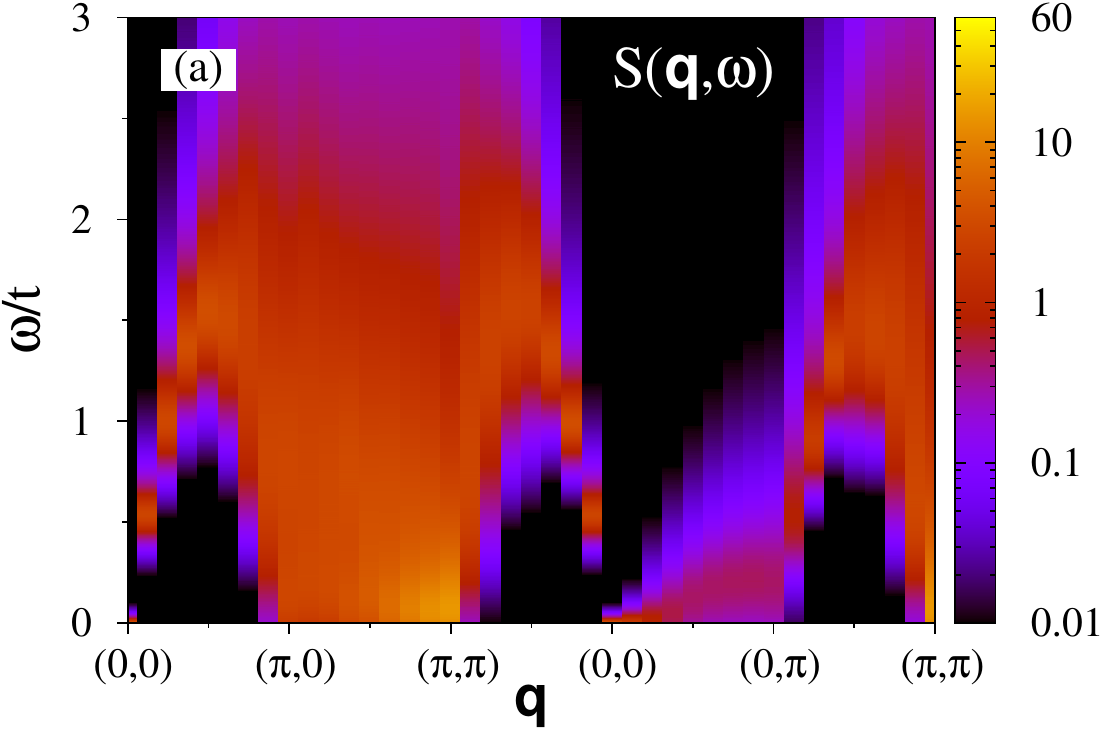}\\
\includegraphics*[width=0.43\textwidth]{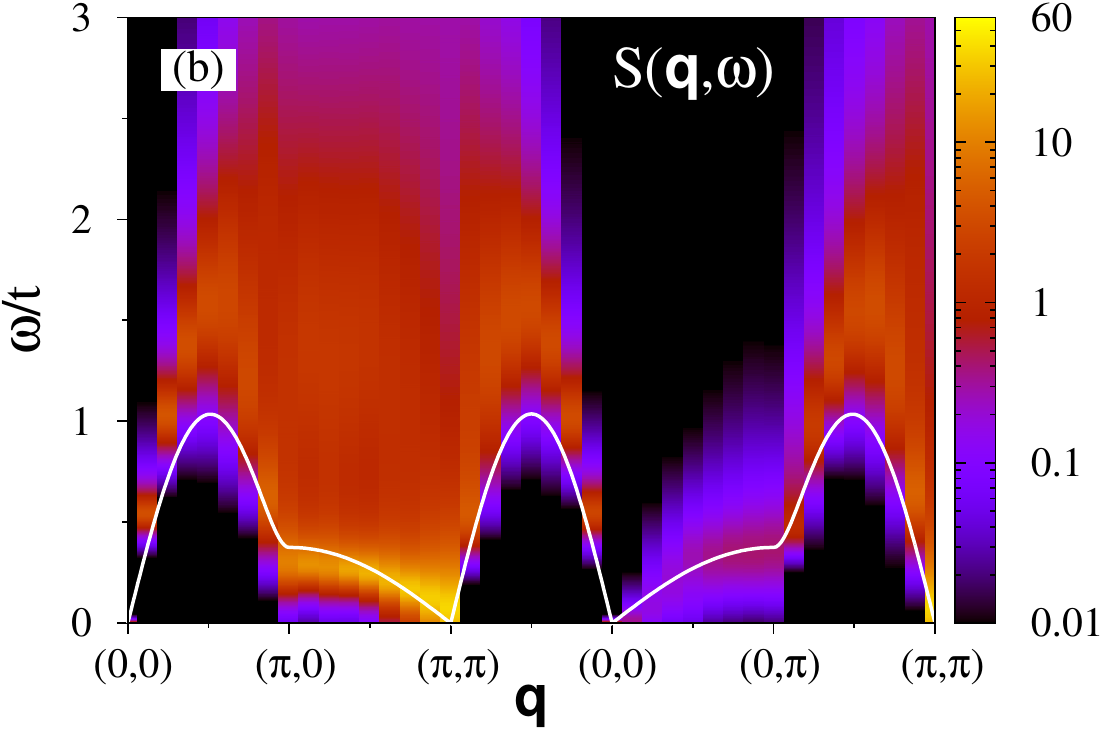}
\end{center}
\caption {(Color online) Dynamical spin structure factor   $S({\pmb q},\omega)$
obtained on the $16\times 16$ lattice with $t_{\perp}/t=0.2$
at: (a) $T=t/10$ and (b) $T=t/20$. Solid line in the lower
plot gives a LSWT fit Eq.~(\ref{magnon}) with $J_{\perp}/J=0.06$ and $J'/J_{\perp}=0.2$.
}
\label{Sqw2}
\end{figure}

\begin{figure}[t!]
\begin{center}
\includegraphics*[width=0.43\textwidth]{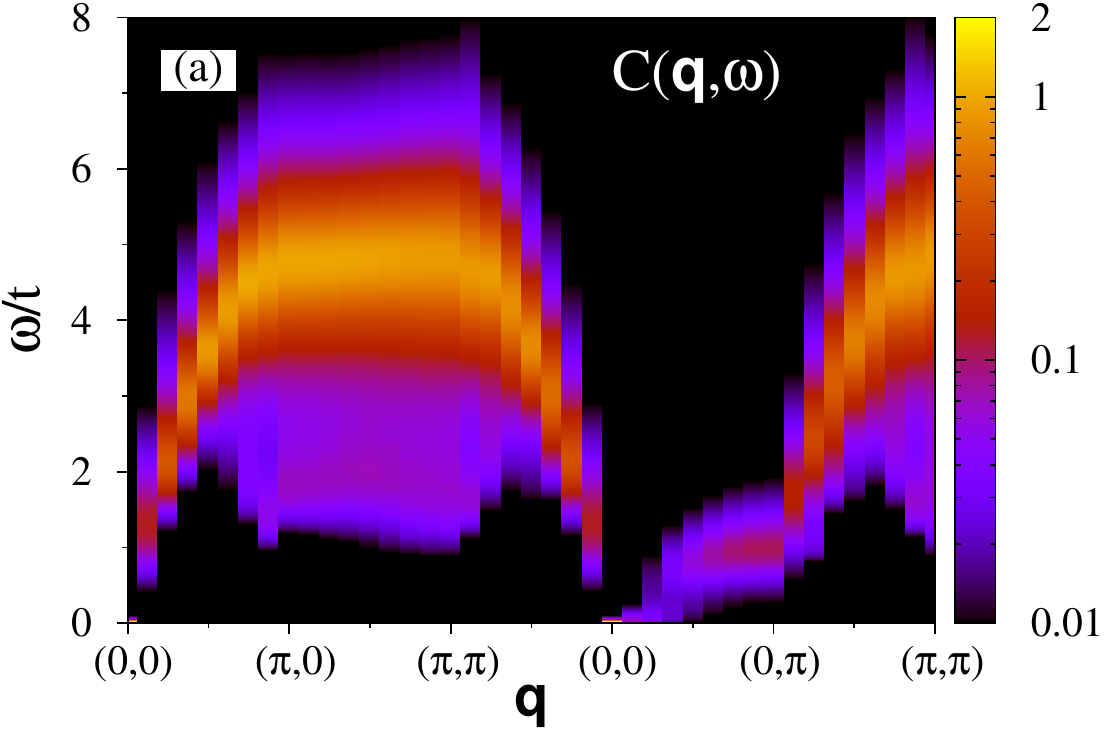}\\
\includegraphics*[width=0.43\textwidth]{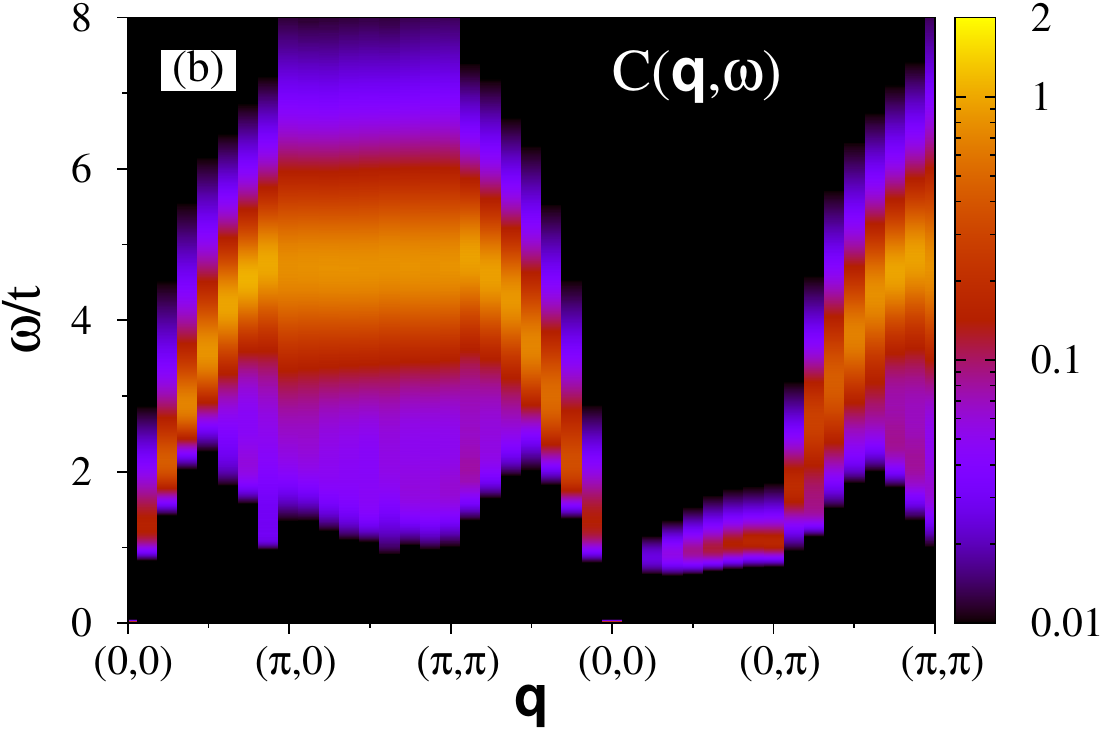}
\end{center}
\caption {(Color online) Dynamical charge structure factor   $C({\pmb q},\omega)$
obtained on the $16\times 16$ lattice with $t_{\perp}/t=0.2$
at: (a) $T=t/10$ and (b) $T=t/20$.
}
\label{Cqw2}
\end{figure}

We turn now to spin and charge excitation spectra of weakly coupled Hubbard chains. 
Coupling the chains affects first the low-frequency spin excitations. 
However, due to a small velocity, the spin-wave mode in the small regime of 
$t_{\perp}$ might be resolved only at sufficiently low-$T$. Hence, we begin the discussion by 
showing the data at $t_{\perp}/t=0.2$,\footnote{Assuming the strong coupling relation $J=4t^2/U$ 
with the intrachain (interchain) coupling $J=34$ meV ($J_{\perp}=-1.6$ meV),\cite{Lake05} respectively, 
one finds for KCuF$_3$ $t_{\perp}/t\simeq 0.22$.} 
and $T=t/10$, see Fig.~\ref{Sqw2}(a). In this case one clearly observes:
(i) reduction of $S({\pmb q},\omega)$ around momentum ${\pmb q}=(\pi,0)$ with respect to the 1D regime;
(ii) buildup of the magnetic weight at the AF wave vector ${\pmb q}=(\pi,\pi)$,
and (iii) emergence of a broad incoherent feature along the $(0,0)\to(0,\pi)$ direction.
Upon decreasing temperature down to $T=t/20$, the features (i) and (ii) 
evolve into  a gapless spin-wave mode, see Fig.~\ref{Sqw2}(b).  
It replaces the low-energy part of the two-spinon continuum. 
This should be contrasted with spinon confinement in a two-leg Hubbard~\cite{Hur01} or 
spin $S=1/2$ Heisenberg~\cite{Dagotto92,Shelton96,Greiter02a,Greiter02b} ladder systems with  
a finite interchain coupling leading to a singlet ground state with gapped magnons. 

Spinon confinement is also reflected in the overall loss of spectral weight  of the two-spinon continuum. 
This effect is particularly strong around the AF wave vector 
${\pmb q}=(\pi,\pi)$ where the low-energy magnetic intensity  quickly fades away on moving to 
higher frequencies. 
 
As shown in Fig.~\ref{Sqw2}(b), the LSWT dispersion relation Eq.~(\ref{magnon}) provides quite a good 
description of the low-energy part of the magnon spectrum along the $(\pi,\pi)\to(\pi,0)$ path. 
However, a certain discrepancy emerges at higher energies. Note, that for the chosen fitting parameters 
$J_{\perp}/J=0.06$ and $J'/J_{\perp}=0.2$, the LSWT yields a finite magnetic order parameter 
$\langle S^z\rangle$ (see Table~\ref{J_c}) consistent with the long-range AF order in the system.
Spectral weight is also found along the $(0,0)\to(0,\pi)$ direction. However, its intensity is low and 
vanishes in the  $|{\pmb q}| \to 0$ limit.   
Furthermore, there is a clear anisotropy in the spin-wave velocity. 
The small velocity associated with the interchain dispersion relation, i.e., from $(0,0)\to(0,\pi)$,  
renders thermal effects stronger, and  could provide an explanation  for the broad line shape even in 
the low-energy limit. On the other hand and at higher energies, the magnon can decay into spinons,  
thus providing a damping mechanism of the magnon mode.

The dimensional crossover is also seen in the charge dynamics shown in Fig.~\ref{Cqw2}. 
Indeed, the comparison with the 1D case (see Fig.~\ref{Cqw0}) reveals:  
(i) disappearance of the 1D  2$k_{\rm F}$ feature observed in Fig.~\ref{Cqw0}, and
(ii) emergence of a broad charge mode dispersing along the $(0,0)\to(0,\pi)$ direction.  
As depicted in Fig.~\ref{Cqw2}(b), its low-frequency weight is washed out on decreasing 
temperature down to $T=t/20$. This is consistent with  the localization of charges as appropriate 
for the insulating state.

\begin{figure}[t!]
\begin{center}
\includegraphics*[width=0.43\textwidth]{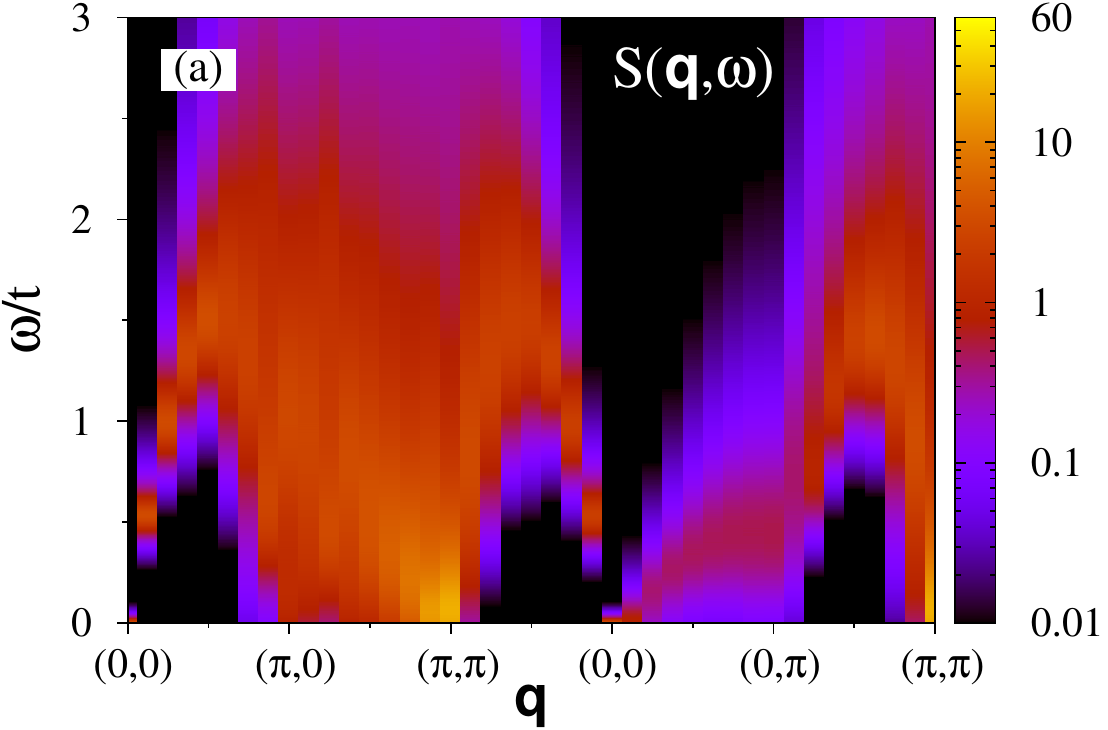}\\
\includegraphics*[width=0.43\textwidth]{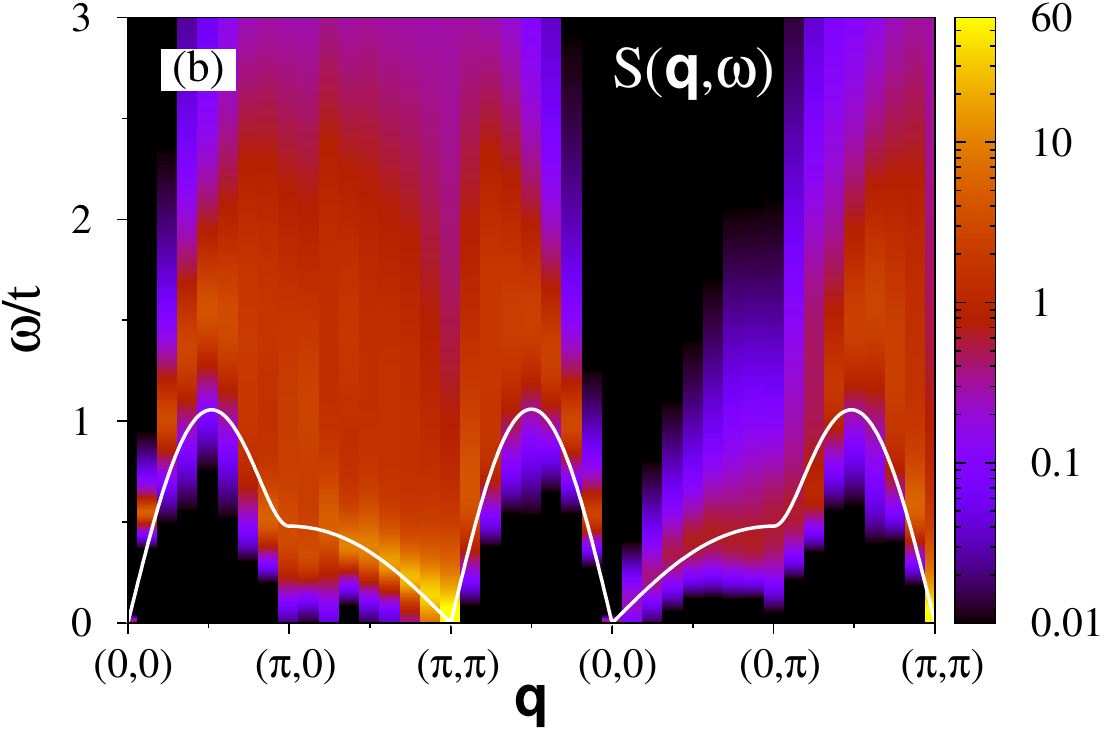}
\end{center}
\caption {(Color online) Same as in Fig.~\ref{Sqw2} but for $t_{\perp}/t=0.3$.
Solid line in the lower plot gives a LSWT fit Eq.~(\ref{magnon}) 
with $J_{\perp}/J=0.1$ and $J'/J_{\perp}=0.2$.
}
\label{Sqw3}
\end{figure}

\begin{figure}[t!]
\begin{center}
\includegraphics*[width=0.43\textwidth]{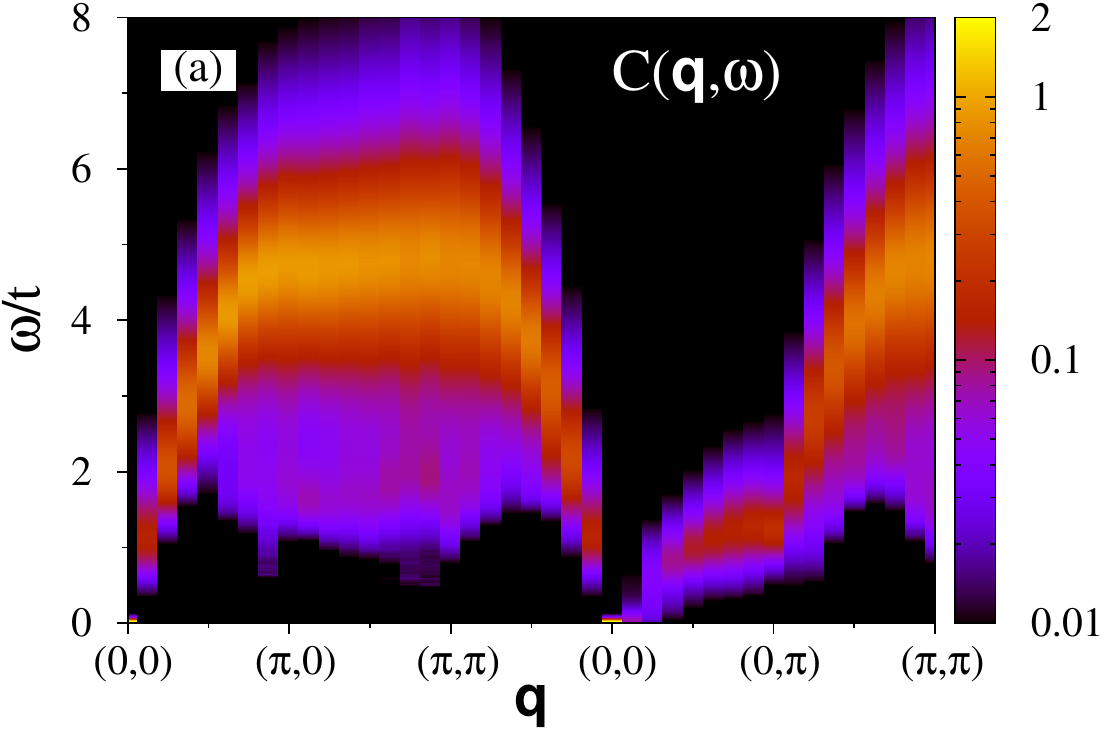}\\
\includegraphics*[width=0.43\textwidth]{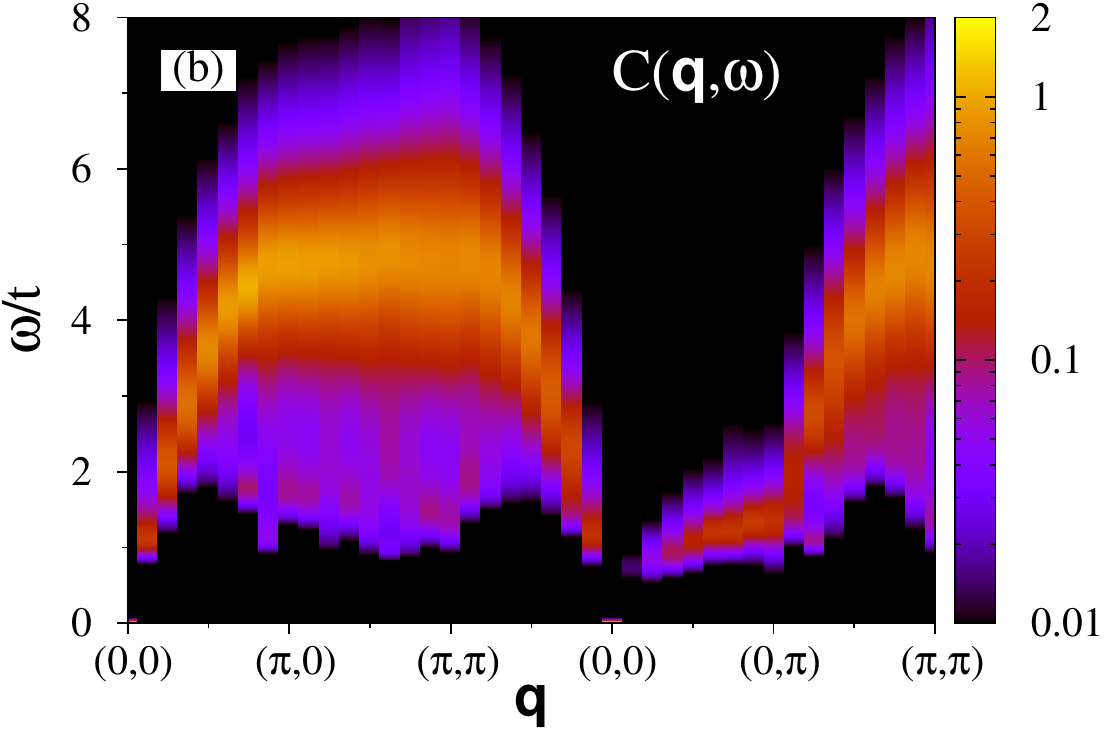}
\end{center}
\caption {(Color online) Same as in Fig.~\ref{Cqw2} but for $t_{\perp}/t=0.3$.
}
\label{Cqw3}
\end{figure}

The inspection of both the dynamical spin, $S({\pmb q},\omega)$,  and  charge, $C({\pmb q},\omega)$, 
structure factors allows us to identify  essentially \emph{two} frequency regimes dominated by 
magnetic excitations of different nature:
(i) low-frequency magnons, and (ii) intermediate-frequency two-spinon excitations.
We interpret this energy-scale separation  as follows. 
Coupling the chains triggers binding of spinons into low-energy spin-waves.  However, 
deconfinement of spinons still occurs above a threshold energy set up by the strength of 
attractive potential between the spinons. 
In proximity to the 1D regime, this potential might be easily overcome by 
thermal fluctuations. This accounts for the observed  transfer  magnetic weight out of the magnon peak 
into the two-spinon continuum.  We illustrate this effect in Fig.~\ref{Sqw2} which compares
$S({\pmb q},\omega)$ for two different temperatures $T=t/10$ and $T=t/20$.

Let us now make a comparison between the spin excitation spectra  $S({\pmb q},\omega)$ 
found at $t_{\perp}/t=0.2$ and $t_{\perp}/t=0.3$. The latter is shown in Fig.~\ref{Sqw3}.
It may be seen that at low temperature $T=t/20$  the spin-wave mode acquires progressively 
increasing damping on going from $(\pi,\pi)$ to $(\pi,0)$, see Fig.~\ref{Sqw3}(b). 
This should be contrasted with $S({\pmb q},\omega)$ at $t_{\perp}/t=0.2$ where the magnon dispersion 
is clearly resolved across the whole $(\pi,\pi)\to(\pi,0)$ path, see Fig.~\ref{Sqw2}(b).   
We ascribe this enhanced decay rate to larger bandwidth of the spin-wave dispersion. 
As a result, magnon excitations reach frequencies high-enough to couple with charge excitations. 
Hence, in contrast to $t_{\perp}/t=0.2$ with a single decay chanel, i.e., into pairs of spinons, 
spin-waves at $t_{\perp}/t=0.3$ might also spontaneously decay into charge excitations. 
Nevertheless, the low-energy part of the magnon dispersion is fairly well accounted within the LSWT, 
see Fig.~\ref{Sqw3}(b). 

Stronger spinon confinement expected for $t_{\perp}/t=0.3$ is consistent with larger amount 
of the magnon weight perpendicular to the chains, i.e., along the $(0,0)\to(0,\pi)$ path. 
Accumulation of spectral weight along the same $(0,0)\to(0,\pi)$ direction is also observed in the charge 
excitation spectrum $C({\pmb q},\omega)$, see  Fig.~\ref{Cqw3}. It reflects development of 
charge correlations between the chains. In contrast, there are no zero-frequency charge excitations 
in the low-temperature regime $T=t/20$ typical of the insulating phase, see Fig.~\ref{Cqw3}(b). 
However, this finite-energy charge mode might be considered as a precursor feature of 
electron deconfinement, i.e., possibility of the charge transfer across the chains in 
the dimensional-crossover-driven Mott transition.~\cite{Raczkowski12}
Finally, a weak momentum dependence of charge excitations may be resolved along 
the $(\pi,0)\to(\pi,\pi)$ path.

\section{Conclusions}
\label{Conclusions}

We have systematically studied  the evolution of spin and charge degrees of freedom upon  coupling 
the 1D Hubbard chains with frustrating hopping matrix elements.   From the technical point of view, 
we tackled  the problem  with the numerically exact finite-temperature auxiliary-field QMC algorithm  
on lattice sizes up to $20\times 20$.  At the considered value of $U/t=3$ and up to 
$t_{\perp} /t = 0.35$,   the sign problem turned out to be manageable down to $T = t/20$.  
In this parameter regime, the low-temperature dynamical charge structure factor shows that the system 
remains insulating.

The 1D limit is very well understood.~\cite{Giamarchi_book} The relevance of umklapp processes 
at half-filling opens a charge gap and results in critical equal-time spin-spin correlations. 
The spin dynamical structure factor  exhibits the two-spinon continuum with  gapless excitations 
at long wavelengths and at ${\pmb q} = \pi$.  Upon coupling the chains, our results support  
the  direct onset of a broken-symmetry AF ground state. 
This result implies that spinons will  bind  into magnons as soon as the chains are coupled.

Our numerical results  for the dynamical spin structure factor resolve the  frequency  and temperature
dependence of the transition from deconfined to confined spinons.
In  particular, a transverse spin-wave mode emerges in the low-energy sector of the dynamical 
spin structure factor.
This spin-wave mode is heavily damped since it can decay into pair of spinons present at higher energies.
The overall high-energy features of the dynamical spin structure factor  show clear signatures of
the two-spinon continuum which progressively fade away as the interchain coupling is  enhanced.
The same is valid for the dynamical charge structure factor which shows robust 1D features at 
energy scales beyond the charge gap and up to our largest value of the interchain coupling.

Finally, let us discuss a possible experimental relevance of our results. 
A similar crossover in the nature of spin excitations from dispersive scattering continua to 
sharp magnon modes at the lowest energies has been observed below the ordering temperature $T_N$ 
in a moderately anisotropic \emph{triangular} lattice 
of a $S=1/2$ antiferromagnet Cs$_2$CuCl$_4$.~\cite{Coldea01,Coldea03}
The complex structure  of magnetic excitations  has been ascribed in this case to the proximity 
of Cs$_2$CuCl$_4$ to a fractionalized spin liquid phase which sets up as 
a result of an effective dimensional reduction by the strong geometric 
frustration.~\cite{Kohno07}

The coexistence  of the one- and higher-dimensional transverse spin dynamics operating at 
different frequency scales, has also been resolved in the inelastic neutron scattering 
data on BaCu$_2$Si$_2$O$_7$ and KCuF$_3$, both of them consisting of weakly coupled $S=1/2$ 
chains.~\cite{Zhe00, Zhe01, Lake05}
A detailed comparison of the dynamical spin correlation function $S({\pmb q},\omega)$ for 
the anisotropic $A$-type antiferromagnetic ($A$-AF) phase of KCuF$_3$ should take into 
account a weak \emph{ferromagnetic} superexchange between the AF chains. The latter is 
expected to facilitate the formation of spin-waves by reducing quantum corrections to the 
order parameter in the $A$-AF phase.~\cite{Raczkowski02} 
As such, it might shift a threshold energy above which one recovers the two-spinon continuum 
towards higher energies as compared to the antiferromagnet with solely AF interactions.   

We conclude that the simultaneous observation of low-energy \emph{magnons} and high-energy \emph{spinons} 
is a fingerprint of magnetically ordered phase coexisting with strong quantum fluctuations brought by 
reduced dimensionality.


\begin{acknowledgments}
Discussions with L. Pollet and insightful correspondence with A.~M. Tsvelik are kindly acknowledged.
We thank the LRZ-M\"unich and the J\"ulich Supercomputing center for a generous 
allocation of CPU time and acknowledge support from the DFG grant AS120/8-2 (FOR1346).   
M.~R. is supported by the FP7/ERC Starting Grant No. 306897 ("QUSIMGAS") 
and partially by Polish National Science Center (NCN) under Project No. 2012/04/A/ST3/00331.
\end{acknowledgments}

\bibliographystyle{apsrev4-1}

%

\end{document}